\documentclass[10pt,showpacs,floatfix,letterpaper,amsmath,amsfonts,amssymb,aps,prl,reprint]{revtex4-1}
\usepackage[latin1]{inputenc}
\usepackage{bm}
\usepackage{graphicx}
\usepackage{tensor}
\usepackage{epstopdf}
\usepackage{enumerate}

\newcommand{\op}[1]{\hat{#1}}                                 
\newcommand{\ket}[1]{\lvert #1\rangle}                        
\newcommand{\bra}[1]{\langle #1 \rvert}                       
\newcommand{\abs}[1]{\left\lvert #1 \right\rvert}             
\newcommand{\pr}[1]{\ket{#1}\bra{#1}}                         
\newcommand{\mean}[1]{\langle #1 \rangle}                     

\begin{document}
\title{Avoiding Loopholes with Hybrid Bell-Leggett-Garg Inequalities}
\author{Justin Dressel}
\author{Alexander N. Korotkov}
\affiliation{Department of Electrical Engineering, University of California, Riverside, CA 92521, USA.}

\date{\today}

\begin{abstract}
  By combining the postulates of macrorealism with Bell locality, we derive a qualitatively different hybrid inequality that avoids two loopholes that commonly appear in Leggett-Garg and Bell inequalities.  First, locally invasive measurements can be used, which avoids the ``clumsiness'' Leggett-Garg inequality loophole.  Second, a single experimental ensemble with fixed analyzer settings is sampled, which avoids the ``disjoint sampling'' Bell inequality loophole.  The derived hybrid inequality has the same form as the Clauser-Horne-Shimony-Holt Bell inequality; however, its quantum violation intriguingly requires weak measurements.  A realistic explanation of an observed violation requires either the failure of Bell locality, or a preparation conspiracy of finely tuned and nonlocally correlated noise.  Modern superconducting and optical systems are poised to implement this test.
\end{abstract}


\maketitle

To formally describe the behavior that we expect from the macroscopic world, Leggett and Garg introduced a set of postulates that any theory of macroscopic objects would reasonably obey \cite{Leggett1985,*Leggett2002}.  They dubbed these postulates \emph{macrorealism} (MR) and used them to derive a set of inequalities---now called Leggett-Garg inequalities (LGIs)---that one would expect sequences of measurements on macroscopic objects to satisfy.  These inequalities are formally similar to the Bell inequalities that test the postulates of local realism \cite{Bell1965,Clauser1969}, but involve making multiple measurements on the same object at different points in time.  Quantum theory manifestly violates such LGIs, making them a practical test for the ``quantumness'' of a particular physical system \cite{Lambert2010,*Li2012}.

Though the original inequalities involved noiseless (i.e., projective) detectors, the derivations have been recently generalized to include noisy (i.e., weak) detectors \cite{Ruskov2006,*Jordan2006a,*Williams2008,*Barbieri2009,Wilde2012}.  This generalization has enabled the experimental test of LGIs in superconducting \cite{Palacios-Laloy2010,Groen2013}, optical \cite{Goggin2011,Dressel2011,Xu2011,Waldherr2011,Suzuki2012}, and nuclear magnetic resonance systems \cite{Athalye2011,Souza2011,Knee2012,Katiyar2013}, as well as nitrogen vacancy centers in diamond \cite{George2013}.  See Ref. \cite{Emary2013} for a thorough review of the derivations of generalized LGIs and recent experiments.

A generic shortcoming of these LGIs is that they assume \emph{noninvasive} noisy measurements.  Faced with an LGI violation, a skeptical macrorealist may appeal to hidden invasiveness to explain the violations.  This caveat has been called the ``clumsiness loophole''  \cite{Wilde2012}.  So far, only ``null-result'' measurements have been argued to avoid this loophole \cite{Leggett1985,*Leggett2002,Wilde2012,Knee2012}, since a detector which does not report a result could not classically interact with the system; however, there is still controversy regarding the effectiveness of this strategy \cite{Emary2013}.  As such, this loophole still presents a fundamental obstacle to the interpretation of LGI violations as intrinsic failures of MR.

In a similar vein, a skeptical local realist may discount violations of a typical Bell inequality, such as the Clauser-Horne-Shimony-Holt (CHSH) inequality \cite{Clauser1969}, as an artifact of the ``disjoint sampling loophole'' \cite{Larsson1998}.  This loophole states that since a Bell inequality combines multiple correlators using different analyzer settings of an apparatus, then it is possible that it combines data from distinct and incompatible setting-dependent ensembles.  Thus, a skeptic may argue that a violation merely indicates the incompatibility of the sampled ensembles (e.g., due to setting-dependent coupling efficiencies), rather than the failure of the local realism postulates themselves.

This paper points out the possibility of combining the postulates of MR and Bell locality to avoid both of these common loopholes, leveraging techniques established in Ref.~\cite{Dressel2011}.  Indeed, appending the postulate of Bell locality to MR leads directly to a hybrid Bell-LGI (BLGI) of joint sequential measurements that formally resembles the CHSH Bell inequality.  This hybrid inequality permits locally invasive measurements, so avoids the clumsiness loophole.  It also samples a single experimental ensemble, which avoids the disjoint sampling loophole.

As also shown by Marcovitch and Reznik \cite{Marcovitch2010}, implementing such a CHSH-like correlator using quantum weak measurements on correlated qubits will reproduce the behavior of the standard CHSH correlator.  Thus, the hybrid BLGI derived here can be violated using the same analyzer settings as the CHSH inequality.  Such a violation implies either the failure of Bell locality or a preparation conspiracy that produces nonlocal detector-noise correlations.  We suggest possible implementations of this test that are suitable for current superconducting and optical systems.

\emph{Macrorealism}.---The concept of MR as defined by Leggett and Garg \cite{Leggett1985,*Leggett2002} consists of three key postulates used to derive traditional LGIs:
\begin{enumerate}[(i)]
  \item If an object has several distinguishable states, then at any given time it is in only one of them.
  \item Measuring an object does not disturb its state or its subsequent dynamics.
  \item Measured results are determined causally by prior events.
\end{enumerate}

Generalized LGIs \cite{Ruskov2006,*Jordan2006a,*Williams2008,*Barbieri2009,Wilde2012} that use noisy detectors require an additional postulate:
\begin{enumerate}[(iv)]
  \item Noisy detectors produce results that are correlated with the object state on average.
\end{enumerate}
This postulate implies the following: if an object has a (potentially hidden) physical state, $\zeta$, that determines a property, $A(\zeta)$, and if a detector (including its local environment) has a fluctuating physical state, $\xi$, then that detector will output a noisy signal, $\alpha(\xi)$, with a probability, $P_A(\xi | \zeta)$, such that the following sum rule is satisfied for every $\zeta$:
\begin{align}\label{eq:noiseaverage}
  \textstyle{\sum_\xi}\, \alpha(\xi)\, P_A(\xi | \zeta) = A(\zeta).
\end{align}
The $\zeta$-dependent distribution $P_A(\xi | \zeta)$ arises from the coupling of the detector to the object and ensures that the correct property value $A(\zeta)$ is properly recovered on average.

\emph{Bell locality}.---To improve upon existing generalized LGIs, it is desirable to remove the postulate (ii) of measurement noninvasiveness.  To accomplish this goal, this postulate can be substituted with the weaker assumption of Bell-locality \cite{Bell1965,Clauser1969}:
\begin{enumerate}[(ii')]
  \item A measurement performed on one object of a spacelike-separated pair cannot disturb measurements made on the second object.
\end{enumerate}
Remote correlations between two separated objects can still exist due to a common joint state $\zeta$ prepared according to some distribution $P(\zeta)$ in the past lightcones of both objects.  However, the detector states $\xi$ are local, and so can become correlated only by coupling to $\zeta$.

\begin{figure}[t]
  \begin{center}
    \includegraphics[width=0.85\columnwidth]{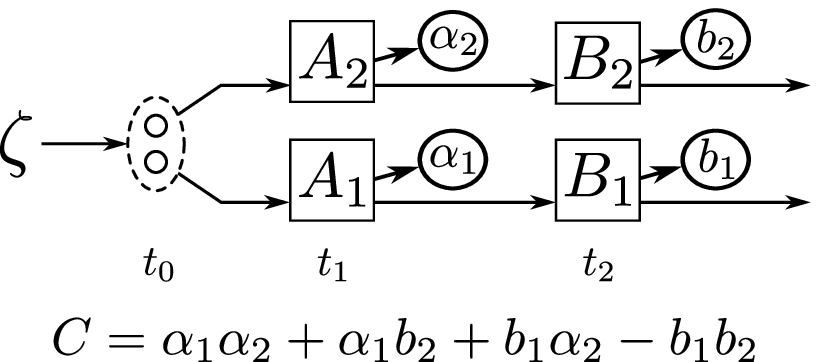}
  \end{center}
  \caption{Schematic for the hybrid Bell-LGI using two pairs of sequential measurements for bounded properties.  The measurements of $A_1$ and $A_2$ have noisy signals $\alpha_1$ and $\alpha_2$ that average to the range $[-1,1]$, while the remaining measurements of $B_1$ and $B_2$ have signals $b_1$ and $b_2$ constrained to the range $[-1,1]$.  The correlator $C$ is averaged over realizations $\zeta$ of the joint preparation $P(\zeta)$.  Postulating Bell locality and MR bounds this correlation to $|\mean{C}| \leq 2$.}
  \label{fig:leggett4}
\end{figure}

\emph{Hybrid Inequality}.---Consider the schematic illustrated in Fig.~\ref{fig:leggett4}.  At time $t_0$ a correlated pair of objects with the joint state $\zeta$ is sampled from an ensemble with the distribution $P(\zeta)$. (We will later consider two qubits, though this derivation is general.) At time $t_1 > t_0$ each object ($k=1,2$) is coupled to a detector with a noisy signal $\alpha_k$ that is calibrated to measure the bounded property $A_k(\zeta)\in[-1,1]$ on average.  The noisy signal $\alpha_k$ generally has an expanded range of values that can lie outside the range $[-1,1]$; however, for each $\zeta$ the realizations of the output signal average to the correct bounded value by assumption (iv).

At time $t_2 > t_1$ each object is then coupled to a second detector with a signal $b_k$ that correlates to a similarly bounded property $B_k(\zeta)\in[-1,1]$.  In contrast to the detectors for $A_k$, each signal $b_k$ is not assumed to be precisely calibrated on average, but is assumed to have the same range as $B_k$.  The reason for this restriction will become clear momentarily.

To obtain an inequality from the four measured signals, we consider the following correlator
\begin{align}\label{eq:corr}
  C = \alpha_1 \alpha_2 + \alpha_1 b_2 + b_1 \alpha_2 - b_1 b_2,
\end{align}
which formally resembles the CHSH correlator \cite{Clauser1969}.  However, the experimenter averages the entire correlator $C$ with every realization of a \emph{single} experimental configuration, in contrast to the Bell case that independently averages each term in $C$ using distinct configurations.  By considering only a single configuration, this BLGI correlator avoids the disjoint sampling loophole \cite{Larsson1998}.

The expanded ranges of the noisy signals $\alpha_k$ generally produce a similarly expanded range for the correlator $C$ for each object pair.  Nevertheless, averaging $C$ over $\xi_{A_k}$ and $\xi_{B_k}$ still produces
\begin{align}\label{eq:corrav}
  \mean{C} &= \textstyle{\sum_\zeta \sum_{\substack{\xi_{A_1},\xi_{B_1} \\ \xi_{A_2},\xi_{B_2}}}}\, C\, P(\xi_{A_1},\xi_{B_1} | \zeta)P(\xi_{A_2},\xi_{B_2}|\zeta)P(\zeta), \nonumber \\
  &= \textstyle{\sum_\zeta} \, \big[A_1(\zeta) A_2(\zeta) + A_1(\zeta) \tilde{B}_2(\zeta) \nonumber \\
    &\qquad + \tilde{B}_1(\zeta) A_2(\zeta) - \tilde{B}_1(\zeta) \tilde{B}_2(\zeta)\big]\, P(\zeta),
\end{align}
with $A_k(\zeta) = \sum_{\xi_{A_k},\xi_{B_k}} \alpha_k(\xi_{A_k})\, P(\xi_{A_k},\xi_{B_k}|\zeta)$ and $\tilde{B}_k(\zeta) = \sum_{\xi_{A_k},\xi_{B_k}} b_k(\xi_{B_k})\, P(\xi_{A_k},\xi_{B_k}|\zeta)$, since postulate (ii') causes the joint distribution of the detector states to factor: $P(\xi_{A_1},\xi_{B_1},\xi_{A_2},\xi_{B_2}|\zeta) = P(\xi_{A_1},\xi_{B_1}|\zeta)P(\xi_{A_2},\xi_{B_2}|\zeta)$.

From the postulates (i), (iii), and (iv), the averages $A_k(\zeta)$ are bounded to the range $[-1,1]$.  Similarly, the averages $\tilde{B}_k(\zeta)$ lie in the range $[-1,1]$ since the signals $b_k(\xi_{B_k})$ are themselves bounded.  Therefore, for each $\zeta$ the sum of the bounded averages in Eq.~\eqref{eq:corrav} must itself be bounded by $[-2,2]$.  Averaging this bounded result with $P(\zeta)$ produces
\begin{align}\label{eq:chsh}
  -2 \leq \mean{C} \leq 2,
\end{align}
in complete analogy to the traditional CHSH inequality.

Importantly, the joint probability $P(\xi_{A_k},\xi_{B_k}|\zeta) = P(\xi_{A_k}|\zeta)P(\xi_{B_k}|\zeta,\xi_{A_k})$ for each arm $k$ admits the dependence of the $B_k$ measurement on an invasive $A_k$ measurement.  Despite any randomization of $b_k(\xi_{B_k})$ from this local invasiveness, however, the perturbed averages $\tilde{B}_k(\zeta)$ must still lie in the range $[-1,1]$.  This allowance for locally invasive measurements in the BLGI avoids the clumsiness loophole.  Note that if $b_k(\xi_{B_k})$ also had an expanded range, as assumed in Ref.~\cite{Marcovitch2010}, then Eq.~\eqref{eq:chsh} would be guaranteed only for noninvasive $A_k$, so the clumsiness loophole would remain.

There are two notable ways that our derivation of the BLGI in Eq.~\eqref{eq:chsh} could fail.  First, the assumption (ii') of Bell locality could fail, either by itself or as a consequence of the realism assumption (i) failing \cite{Norsen2006}.  Second, the noisy detector assumption (iv) could fail due to hidden preparation noise $\xi_P$ (i.e., not included in the object state $\zeta$) that systematically affects the detector output in both arms.  In this case, the detector distributions become noise-dependent $P_A(\xi|\zeta) \to P_A(\xi|\zeta,\xi_P)$ such that Eq.~\eqref{eq:noiseaverage} is satisfied only after additionally averaging over $\xi_P$.  Such joint noise-dependence would prevent the detector distributions from factoring for each $\zeta$ in Eq.~\eqref{eq:corrav}, which spoils the inequality.  However, such a systematic bias due to shared preparation noise can be checked during detector calibration by deliberately preparing a variety of uncorrelated distributions $P(\zeta)$ and looking for spurious cross-correlations caused by such hidden preparation noise.  Hence, the failure of assumption (iv) additionally requires a preparation-conspiracy where every implementable calibration check is apparently free from anomalous noise-correlations.

For an implementation with low pair-collection efficiency (e.g., using optical photodetectors) a related detection loophole also arises.  Specifically, if the ensemble of pairs is unfairly sampled, then the averaging property of Eq.~\eqref{eq:noiseaverage} may not be satisfied, which also causes the failure of assumption (iv).  Hence, the fair sampling assumption will still be needed unless the collection is efficient (e.g., with superconducting qubits).

\emph{Quantum violations}.---The quantum mechanical equivalent of a noisy MR measurement is a weak measurement \cite{Dressel2010,*Dressel2012b,*Dressel2013b,*Dressel2013}, as emphasized in Ref.~\cite{Dressel2011}.  An implementation of Fig.~\ref{fig:leggett4} that uses a particular class of weak measurements was considered in Marcovitch and Reznik~\cite{Marcovitch2010}.  Their analysis shows that a correlation $\mean{C}$ of the CHSH form as in Eq.~\eqref{eq:corr} can saturate the standard quantum bound of $2\sqrt{2}$ in the limit of ideally weak measurements, violating the BLGI just derived in Eq.~\eqref{eq:chsh}.  (This CHSH-like bound of $2$ for the BLGI is assumed without derivation in Ref.~\cite{Marcovitch2010}.)

More generally, any implementation of Fig.~\ref{fig:leggett4} using sufficiently weak qubit measurements can saturate the standard quantum CHSH bound of $|\mean{C}| \leq 2\sqrt{2}$.  This fact can be understood in a simple way: a weak measurement leaves the initial state nearly unperturbed, so all four measurements of $A_k$,$B_k$ probe approximately the same quantum state.  Thus, they will exhibit the same correlations that occur in the standard CHSH inequality \cite{Clauser1969}.  For quantum nondemolition (QND) measurements \cite{Braginski1992}, the difference between the weakly measured BLGI correlator and the traditional CHSH correlator will depend only on the effective \emph{ensemble-dephasing} (i.e., decoherence) that is induced by the $A_k$ measurements.

To see this, note that the maximum CHSH value of $\mean{C} = 2\sqrt{2}$ can be obtained from an entangled Bell state preparation $\ket{\Psi} = (\ket{0,0} + \ket{1,1})/\sqrt{2}$.  The standard bases producing this violation are given by
\begin{subequations}
\begin{align}
  \ket{0}_{D} &= \cos(\phi_D/2)\ket{0} + \sin(\phi_D/2)\ket{1}, \\
  \ket{1}_{D} &= -\sin(\phi_D/2)\ket{0} + \cos(\phi_D/2)\ket{1},
\end{align}
\end{subequations}
with index $D=A_1,A_2,B_1,B_2$ and associated angles $\phi_{B_1} = 0$, $\phi_{B_2} = 3\pi/4$, $\phi_{A_1} = \pi/2$, and $\phi_{A_2} = \pi/4$.  If the $A_k$ measurements are QND and induce dephasing factors $\Xi_k \in [0,1]$ (i.e., the reduced-state coherences update as $\rho_{01} \to \Xi_k\, \rho_{01}$), then averaging the correlator in Eq.~\eqref{eq:corr} with these basis choices produces
\begin{align}\label{eq:blgicorr}
  \mean{C} = (1 + \Xi_1)(1 + \Xi_2)/\sqrt{2}.
\end{align}
For projective measurements, $\Xi_k \to 0$, so $\mean{C} \to
1/\sqrt{2}$, while for ideally weak measurements, $\Xi_k \to 1$, so
$\mean{C} \to 2\sqrt{2}$.  The specific form of the dephasing does
not matter for this general result.  Moreover, if the $B_k$
measurements have lower visibility $v \in [0,1]$ (e.g., due to
misidentification errors), then the effective dephasing is simply
enhanced $\Xi_k \to v\, \Xi_k$.  A violation of Eq.~\eqref{eq:chsh} will
occur whenever $v\, \Xi_k > -1 + 2^{3/4} \approx 0.68$, which provides
a practical lower bound for the visibility $v$.

An explanation of these quantum-predicted BLGI violations as a failure of assumption (iv) due to noise correlations requires not only preparation conspiracy, but also carefully tuned noise.  That is, the noise produces results that can also be measured from the same preparation using the standard CHSH protocol (e.g., by using the $B_k$ measurements alone and varying the analyzer settings).  Such fine-tuning cannot be easily attributed to random environmental fluctuations during the preparation.

\emph{Gaussian meter}.---For specificity, consider a Gaussian measurement of $A_k$ with variance $\sigma_k^2$.  Implementations of such a Gaussian measurement have been discussed for optical \cite{Aharonov1988,*Ritchie1991,Wiseman2009}, quantum dot \cite{Ruskov2006,*Jordan2006a,*Williams2008,Korotkov2001,*Korotkov2011}, and superconducting \cite{Korotkov2001,*Korotkov2011,Wiseman2009,Palacios-Laloy2010,Vijay2012,Hatridge2013} systems.  For Gaussians that do not appreciably overlap, $\sigma_k \ll 1$, the measurement is ideal (strong), and the measured detector output perfectly correlates with $A_k$.  Conversely, for overlapping Gaussians, $\sigma_k \gg 1$, the measurement is noisy (weak), and the output poorly correlates with $A_k$.  In the limit that $\sigma_k \to \infty$ the measurement becomes ideally weak.

\begin{figure}[t]
  \begin{center}
    \includegraphics[width=0.85\columnwidth]{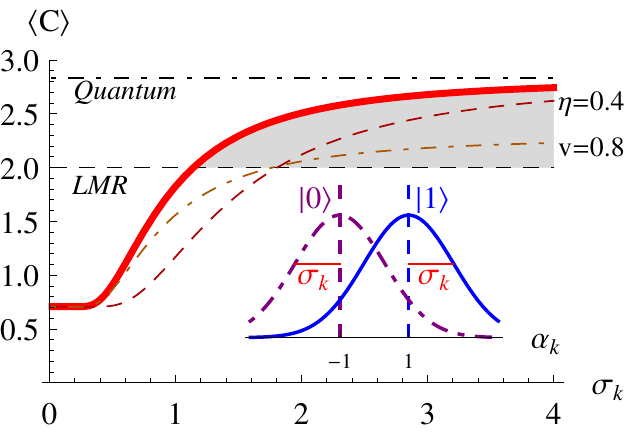}
  \end{center}
  \caption{(color online) Bell-LGI correlator $\mean{C}$, where the
$A_k$ eigenvalues $\pm 1$ are measured with signals $\alpha_k$
having Gaussian probability distributions (inset) with standard
deviations $\sigma_k$.  In the projective limit $\sigma_k \to 0$ the
correlator converges to $\mean{C} \to 1/\sqrt{2}$, while in the weak
limit $\sigma_k \to \infty$ it violates the local-macrorealistic
(LMR) bound of $2$ and converges to the quantum bound of
$2\sqrt{2}$.  The LMR bound is still violated for moderate $B_k$
visibility $v\in[0,1]$, while the efficiency $\eta\in[0,1]$ simply
scales the $\sigma_k$ required to see the violations.}
  \label{fig:gaussc}
\end{figure}

A quantum-limited Gaussian qubit measurement corresponds to the following partial projection operator
\begin{align}\label{eq:gaussmeas}
  \op{M}_{\alpha_k} &= \frac{e^{-(\alpha_k - 1)/4\sigma_k^2}\pr{0} + e^{-(\alpha_k + 1)/4\sigma_k^2}\pr{1}}{\left(2\pi\sigma_k^2\right)^{1/4}},
\end{align}
where the output signal $\alpha_k$ has a mean centered on each qubit eigenvalue of $\pm 1$ corresponding to eigenstates $\ket{0}_{A_k}$ and $\ket{1}_{A_k}$ in the basis of $A_k$ (as shown in the inset of Fig.~\ref{fig:gaussc}).  This partial projection leads to dephasing $\Xi_k = e^{-1/2\sigma_k^2}$. For a detector with efficiency $\eta \in [0,1]$, there are additional imperfections that result in a faster dephasing of $\Xi_k = e^{-1/2\sigma_k^2\eta}$ \cite{Korotkov2001,Wiseman2009}.

In the ideal case with $\eta = 1$, the joint probability of the four measurements indicated in Fig.~\ref{fig:leggett4} is
\begin{align}\label{eq:gaussp}
  P(\alpha_1,\alpha_2,b_1,b_2 | \Psi) &= \abs{ \bra{b_1,b_2}\op{M}_{\alpha_1}\!\otimes\!\op{M}_{\alpha_2}\ket{\Psi} }^2
\end{align}
with the joint state $\ket{\Psi}$ corresponding to the Bell-state
preparation and $\ket{b_1,b_2} =
\ket{b_1}_{B_1}\otimes\ket{b_2}_{B_2}$ corresponding to projective
measurements in the basis for $B_k$ with eigenvalues $b_k = \pm 1$.
Note that misidentification errors in $B_k$ lower the measurement
fidelity by decreasing the signal visibility: $b_k = \pm v$, with
$v\in[0,1]$.

Averaging the correlator of Eq.~\eqref{eq:corr} with the
distribution of Eq.~\eqref{eq:gaussp} produces
Eq.~\eqref{eq:blgicorr} with $\Xi_k = e^{-1/2\sigma_k}$.  Including
inefficiency and visibility factors $\eta,v\in[0,1]$ for $A_k$ and
$B_k$ further enhances the dephasing to $\Xi_k = v\,
e^{-1/2\eta\sigma_k}$.  These predictions are illustrated in
Fig.~\ref{fig:gaussc}.  Note that only $v$ alters the obtainable
upper bound.

\emph{Ancilla qubit}.---As an alternative to a Gaussian meter, one can perform an indirect measurement via an ancilla qubit. With such a scheme, the ancilla is entangled with the main qubit and then measured. After the entangling operation the state has the form $(\op{M}_+ \ket{\psi})\ket{+} + (\op{M}_- \ket{\psi})\ket{-}$, where $\ket{\psi}$ is the initial state of the main qubit and $\ket{\pm}$ is the measurement basis for the ancilla. The operators $\op{M}_\pm$ describe the back-action on the main qubit from the measurement. Such indirect ancilla measurements have been implemented in optical \cite{Pryde2004,*Pryde2005,Goggin2011,Dressel2011} and superconducting \cite{Groen2013} systems.

An ideal ancilla measurement is diagonal in the basis
$\ket{0}_{A_k},\ket{1}_{A_k}$ for $A_k$.  If its backaction is
symmetric, then it can be parametrized by a single (deliberately
reduced) visibility parameter $V_k\in[0,1]$:
\begin{align}\label{eq:qubitmeas}
  \op{M}_{k,\pm} &= \sqrt{\frac{1}{2} \pm \frac{V_k}{2}}\,\pr{0} +
  \sqrt{\frac{1}{2} \mp \frac{V_k}{2}}\,\pr{1}.
\end{align}
The rescaled signal satisfying Eq.~\eqref{eq:noiseaverage} is then
$\alpha_{k,\pm} = \pm 1/V_k$
\cite{Dressel2010,*Dressel2012b,*Dressel2013b,*Dressel2013}.  The
second moment of this signal is $V_k^{-2}$, so the signal variance
for each definite qubit state is $\sigma_k^2 = V_k^{-2} - 1$, which
vanishes for projective measurements with $V_k=1$.  The
ensemble-dephasing due to the ancilla measurement is $\Xi_k =
(1-V_k^{2})^{1/2} = \sigma_k(1 + \sigma_k^2)^{-1/2}$, which is
not exponential (in contrast to the Gaussian case).

\begin{figure}[t]
  \begin{center}
    \includegraphics[width=0.85\columnwidth]{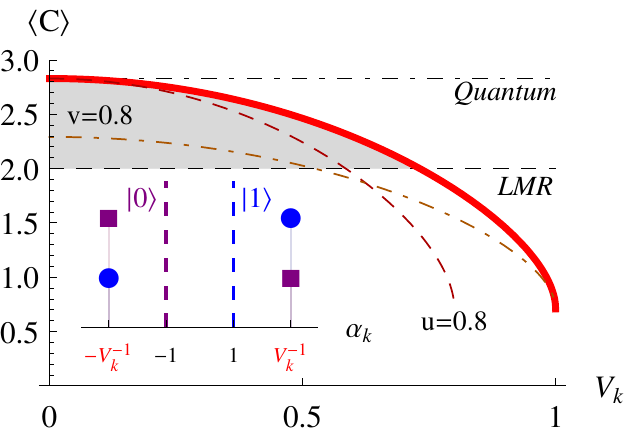}
  \end{center}
  \caption{(color online) Bell-LGI correlator $\mean{C}$, where
each $A_k$ is measured with an ancilla qubit, yielding the signal $\alpha_k =
\pm 1/V_k$ with a discrete probability distribution
(inset as a histogram for comparison with the Gaussian case inset in
Fig.~\ref{fig:gaussc}).  The weak limit as the visibilities $V_k$
approach zero violates the LMR bound of $2$ and converges to the
quantum bound of $2\sqrt{2}$.  The visibility $v\in[0,1]$ for $B_k$
decreases $\mean{C}$ similarly to Fig.~\ref{fig:gaussc}, while the
visibility $u\in[0,1]$ for the ancilla measurement scales the
horizontal axis, $V_k \leq u$.}
  \label{fig:qubitc}
\end{figure}

Replacing the operators $\op{M}_{\alpha_k}$ in Eq.~\eqref{eq:gaussp}
with $\op{M}_{k,\pm}$ from Eq.~\eqref{eq:qubitmeas} produces the
joint probability.  Averaging the correlator in Eq.~\eqref{eq:corr}
with the signal $\alpha_{k,\pm}=\pm 1/V_k$ also produces
Eq.~\eqref{eq:blgicorr} with dephasing $\Xi_k = (1-V_k^{2})^{1/2}$,
violating the BLGI in Eq.~\eqref{eq:chsh}. Introducing
misidentification errors for ancilla measurements and $B_k$ produces
the corresponding visibilities $u,v\in[0,1]$, which further enhance
the effective dephasing in Eq.~\eqref{eq:blgicorr} to $\Xi_k =
v\,[1-(V_k/u)^{2}]^{1/2}$ (here $V_k\leq u$ is the total
visibility). These predictions are illustrated in
Fig.~\ref{fig:qubitc}. As with the Gaussian case, only $v$ reduces
the obtainable upper bound.

\emph{Conclusion}.---The hybrid Bell-LGI derived in this paper formally resembles the CHSH Bell inequality, but combines the postulates of macrorealism and Bell locality.  The derivation avoids both the disjoint sampling loophole of the standard CHSH inequality and the clumsiness loophole of generalized LGIs.  The quantum violation of the Bell-LGI requires weak measurements.  A realistic explanation of these quantum predictions requires the failure of Bell locality or the presence of finely tuned noise-correlations.  Modern superconducting and optical systems are primed to test for these violations.

\emph{Acknowledgments}.---JD thanks Curtis J. Broadbent, David J. Starling, Eyob A. Sete, and Andrew N. Jordan for stimulating discussions.  The research was funded by the Office of the Director of National Intelligence (ODNI), Intelligence Advanced Research Projects Activity (IARPA), through Army Research Office (ARO) Grant No. W911NF-10-1-0334. All statements of fact, opinion, or conclusions contained herein are those of the authors and should not be construed as representing the official views or policies of IARPA, the ODNI, or the U.S. Government. We also acknowledge support from ARO MURI Grant No. W911NF-11-1-0268.

%
\end{document}